# Controllable tunability of a Chern number within the electronic-nuclear spin system in diamond

**Authors:** Junghyun Lee[1,2,†], Keigo Arai[3,4,†], Huiliang Zhang[3,5], Mark J. H. Ku[3,5], and Ronald L. Walsworth[3,5,6,*]

**Affiliations:**

[1]Department of Physics, Massachusetts Institute of Technology, Cambridge, Massachusetts 02139, USA.

[2]Center for Quantum Information, Korea Institute of Science and Technology, Seoul 02792, Republic of Korea.

[3]Harvard-Smithsonian Center for Astrophysics, Cambridge, Massachusetts 02138, USA.

[4]Department of Electrical and Electronic Engineering, Tokyo Institute of Technology, Tokyo 152-8552 Japan

[5]Department of Physics, Harvard University, Cambridge, Massachusetts 02138, USA.

[6]Center for Brain Science, Harvard University, Cambridge, Massachusetts 02138, USA.

†These authors contributed equally to this work.
*Correspondence to: walsworth@umd.edu

**Abstract:**

Chern numbers are gaining traction as they characterize topological phases in various physical systems. However, the resilience of the system topology to external perturbations makes it challenging to experimentally investigate transitions between different phases. In this study, we demonstrate the transitions of Chern number from 0 to 3, synthesized in an electronic-nuclear spin system associated with the nitrogen-vacancy (NV) centre in diamond. The Chern number is characterized by the number of degeneracies enclosed in a control Hamiltonian parameter sphere. The topological transitions between different phases are depicted by varying the radius and offset of the sphere. We show that the measured topological phase diagram is not only consistent with the numerical calculations but can also be mapped onto an interacting three-qubit system. The NV system may also allow access to even higher Chern numbers, which can be applied to exploring exotic topology or topological quantum information.

**Main text:**

**Introduction**

Currently, extensive research is being conducted on the Chern number[1], which is defined as the integral of the Berry curvature[2–6], and on the application of its robust topological properties to quantum metrology[7], next-generation electronics[8], spintronics [9], and quantum computation[10–13]. Particularly, exploration of the Chern number to its higher values and investigating the transitions between them are of significant interest[14–16]. For example, high Chern number phases in the quantum anomalous Hall insulators are a candidate platform for next-generation low-power-consumption electronics because the contact resistance between the normal metal electrodes and chiral edge channels drops as the Chern number increases[17–19]. To characterize this scaling experimentally, it is necessary to vary the Chern number without changing the material properties.

Despite an advanced theoretical foundation, Chern numbers greater than one have been scarcely observed experimentally in condensed matter systems until recently, such as multilayer-graphene boron-nitride interfaces with field-tuneable superlattice flat bands[14,20] and undoped multilayers of topological insulator under alternating magnetic fields[8,21]. Moreover, although the controllability of the Chern number will add tremendous value to the abovementioned practical applications, it is even more challenging to transit across the topological phases. Two major challenges hinder their experimental investigation, namely, continuous tuning of the properties in materials and the direct detection of the topological invariant of the multi-fold degenerate points in condensed matter systems.

An alternative approach to simulating Chern numbers involves the use of two-level systems in various qubit platforms[22–26], including superconducting qubits[27,28], ultracold atoms[29–32], and nitrogen vacancy (NV) centres in diamond[33]. These platforms with large experimental degrees of freedom can be employed as powerful quantum simulators for simulating complex and dynamic Hamiltonian models in condensed matter systems, which are usually difficult to access or even inaccessible. Notably, researchers have used a single NV centre to explore 2D synthetic quantum Hall physics[34] and a synthetic monopole source in the Kalb-Ramond field [35].

According to Gritsev et al.[36], the Chern number can be measured as an integral of the deviation in the qubit Bloch vector from the hemispherical trajectory of a time-varying Larmor vector owing to a nonadiabatic response (Fig. 1a). Here, we assume that the qubit system can be described by the generic Hamiltonian $\widehat{H}(t) = \widehat{H}_S + \widehat{H}_C(t)$, where $\widehat{H}_S$ is the static internal Hamiltonian that defines the degeneracy points, and $\widehat{H}_C(t)$ is the time-dependent control Hamiltonian. The control Hamiltonian adopts the form $\widehat{H}_C(t) = \hbar \vec{H}(t) \cdot \vec{\sigma}/2$, where $\hbar$ is the reduced Planck's constant, $\vec{H}(t)$ is the time-varying Larmor vector in a three-dimensional Hamiltonian parameter space labelled as $(H_x, H_y, H_z)$, and $\vec{\sigma} = (\sigma^x, \sigma^y, \sigma^z)$ are the Pauli matrices. The Larmor vector is chosen to sweep a hemispherical trajectory from the north pole to the south pole with a radius $H_r$ and introduce an offset $H_0$ along the z-axis from the origin:

$$\vec{H}(t) = (H_r \sin\theta(t) \cos\phi(t), H_r \sin\theta(t) \sin\phi(t), H_r \cos\theta(t) + H_0), \quad (1)$$

where $\theta$ is the time-varying polar angle and $\phi$ is the azimuthal angle fixed at 0, without the loss of generality. When the Larmor vector traverses this trajectory at a finite speed, the qubit's Bloch vector $\langle \vec{\sigma} \rangle$ follows the Larmor vector $\vec{H}(t)$, but with a small deviation along the $\phi$ direction at each polar angle location owing to a nonadiabatic response [2–4,37]. For the first-order approximation, this deviation is related to the $\phi$ component of the Berry curvature $F_\phi$ through the following linear relation:

$$F_\phi(\theta) = \frac{H_r \sin\theta \langle \sigma^y \rangle}{2v_\theta}, \quad (2)$$

where $\langle\sigma^y\rangle$ is the expectation value of the y component of the Bloch vector and $v_\theta \equiv d\theta/dt$ denotes the angular speed about its polar axis. An integration of this Berry curvature over the polar angle of the trajectory yields the Chern number as follows:

$$C = \int_0^\pi F_\phi(\theta)d\theta \quad (3)$$

The Chern number depends on the number of degeneracy points of the static internal Hamiltonian enclosed in a control Hamiltonian sphere drawn by the Larmor vector. Every degeneracy point can be regarded as a synthetic magnetic monopole. These monopoles produce radial synthetic magnetic fields that exert a torque on the Bloch vector.

In this work, we apply this protocol to experimentally observe the transition of the Chern number from 0 to 3 using three degeneracy points associated with the ground-state energy level of a single NV centre in diamond (Fig. 1b). The NV electronic spin ground-state has three sublevels $|-1\rangle, |0\rangle$, and $|+1\rangle$, out of which only $|-1\rangle$ and $|0\rangle$ are used as a two-level system, represented by $\vec{\sigma}$ in the following measurements. The host nuclear spin $^{14}$N, with a spin quantum number of $I = 1$, induces hyperfine coupling. The internal Hamiltonian takes the form $\hat{H}_0 = \frac{1}{2}\hbar A_\parallel \sigma^z I^z$, where $A_\parallel/2\pi = 2.2$ MHz is the coupling strength of the longitudinal component of the hyperfine interaction and $I_z$ denotes the z component of the nuclear spin. This electronic-nuclear spin system contains three degeneracy points, allowing us to access the topological phases with a Chern number greater than 1. The time-dependent Larmor vector in Eq. (1) is realized through spin-control microwaves that exhibit a time-varying Rabi frequency $\Omega(t) = \Omega_1 \sin\theta(t)$ and detuning of $\Delta(t) = \Delta_1 \cos\theta(t) + \Delta_2$, both measured in units of Hz. Without loss of generality, the azimuthal angle was set to $\phi = 0$. Under this experimental configuration, the Larmor vector can be written as:

$$\vec{H}(t) = (\Omega_1 \sin\theta(t), 0, \Delta_1 \cos\theta(t) + \Delta_2) \quad (4)$$

Fig. 1c illustrates the experimental sequence for measuring the Berry curvature at a certain polar angle. The hemispherical trajectory starts from the north pole $\theta = 0$ at $t = 0$ and ramps along the

$H_y = 0$ meridian with a constant angular velocity until it reaches the south pole $\theta = \pi$ at $t = T_{ramp}$, that is, $\theta(t) = \pi t (T_{ramp})^{-1}$. Throughout this study, the direction of the trajectory was fixed along the north-to-south direction with respect to the points of ground-state degeneracy. A snapshot of $\langle \sigma^y \rangle$ at various polar angle locations was measured by terminating the sweep at time $t = T_{meas}$.

As a measure of the degree of adiabaticity, an adiabaticity parameter[27] was introduced as follows:

$$\alpha \equiv \frac{\Omega_1 T_{ramp}}{2\pi} \tag{5}$$

This measure represents the fractional change in the Larmor vector. Recalling the extra second-order term $O(v^2)$ in the Berry curvature formula (Supplementary Eqn. (2.1)), $\alpha$ affects the accuracy of the measured Chern number. In the nonadiabatic limit ($\alpha \ll 1$), the first-order approximation of the Berry curvature in Eq. (2) breaks down. Subsequently, the effects of higher-order terms contaminate the signal in our measurements. Conversely, in the adiabatic limit ($\alpha \gg 1$), the NV spin remains in the instantaneous ground state; the spin vector is approximately parallel to the direction of the control field, following the meridian. However, the deviation signal $\langle \sigma^y \rangle$ becomes smaller and eventually lies buried in the noise. For the three-level NV system, the appropriate range reflecting an optimum signal-to-noise ratio was found to be $2 \leq \alpha \leq 10$ (see Supplementary Information). The adiabaticity parameter was set to $\alpha = 2$ for the remainder of the work.

**Results**

As a benchmark experiment, we first characterized a case with the expected Chern number of $C = 0$ (Fig. 2a). This case was realized by choosing a small sphere with a normalized radius of $H_r/A_\parallel = 0.2$ and a normalized detuning of $H_0/A_\parallel = 0.23$, which does not contain any of the three degeneracy points. Although these degeneracy points were expected to make the Berry-curvature zero for any $\theta$, numerical simulations based on a time-dependent Schrödinger equation (see

Methods) predicted a deviation from zero. This deviation can be attributed to the nonadiabatic effect, which limits the accuracy of this quasi-static Chern number measurement approach. The measured Berry curvature was consistent with the simulation results, including the nonadiabatic effect. The resulting Chern number, obtained by integrating the Berry curvature over theta, converged to $C = -0.07 \pm 0.04$. Measurement error is evaluated from the photon-shot noise ($1\sigma$).

To observe higher Chern numbers, we then examined cases with one, two, and three enclosed degeneracy points by increasing the radius up to $H_r/A_\parallel = 2.25$ (Fig. 2b-d). The numerical simulations predicted a larger deviation in the path of the Bloch vector with an increase in the number of enclosed degeneracy points, indicating a more aggressive behaviour of the Berry curvature. The measured Berry curvatures for each case agree well with the numerically simulated values. The Chern numbers were measured to be $C = 0.95 \pm 0.35, 2.20 \pm 0.39$, and $2.93 \pm 0.38$. Thus, our results prove that the NV electronic-nuclear spin system can be used as a platform for synthesizing up to three Chern numbers.

Our proposed NV system can further explore the transition between the observed Chern numbers. Fig. 3a presents the topological phase transition along the normalized radius axis ($H_r/A_\parallel \in \{0.25, 2.25\}$) for various normalized offset conditions: $H_0/A_\parallel = 2.0, 1.0, 0.23, 0.0$. In all the cases, we observed a mild phase transition. This dullness can be attributed to the finite $T_2^*$ time and the limited adiabaticity parameter $\alpha$. Additionally, the consistency between the experimentally measured and numerically simulated Chern numbers reflects this nonadiabatic effect within one standard deviation of the measurement error, except for $H_r/A_\parallel \leq 1$. The disagreement within this small-radius region is possibly due to the imperfect calibration of the low Rabi frequencies (see Methods). One notable effect was found in the case of $H_0/A_\parallel = 0.0$, where the number of enclosed degeneracy points was expected to jump from one to three at $H_0/A_\parallel = 1$. However, in the measurements, a sudden depletion of $C$ was observed near 1, and the transition occurred above 1. This shift in the transition point can be attributed to the nonadiabatic response of a qubit when the Larmor vector coincides with the position of the degeneracy points on the z-axis. Fig. 3b presents the transition curves across the normalized offset ($H_0/A_\parallel \in \{0.00, 2.25\}$) for various values of the normalized radius of $H_r/A_\parallel = 0.23, 0.79, 1.36, 2.17$, advocating the diversity of the phase transition pattern. A systematic view of the measurement results could be obtained by mapping the Chern number phase diagram in a two-dimensional parameter space of $H_0/A_\parallel$ and $H_r/A_\parallel$ (Fig. 3c). It can be observed that the Chern number distribution is not mirror-reflected with respect to the $H_0/A_\parallel = 0$ line. This asymmetry occurs due to the time-reversal symmetry breaking of the system[35] when $C \neq 0$, which is caused by the sweeping of the Larmor vector from positive to negative detuning during the measurement, creating a directional dependence of the Lorentzian force-like response of the qubit; thereby, breaking the Chern number transition symmetry (see Supplementary Information).

**Discussion**

We discuss the connection between the NV system and an interacting three-qubit system to reveal the implications of our two-dimensional topological phase diagram. The topological phase diagram presented in this study was constructed by varying the radius and offset of the sphere with respect to the three degeneracy points. As shown in Fig. 4a, varying the radius under fixed inter-degeneracy spacings is topologically similar to varying the inter-degeneracy spacing under a fixed radius. The former was implemented in this work using the NV centre, which is regarded as a single qubit Ising interacting with an additional spin with a high quantum spin number. The latter can be realized by varying the coupling strength $g$ in an interacting symmetric 1D chain multiqubit system using the following Hamiltonian:

$$\hat{H}_{3q} = -\frac{\hbar}{2}\Big(\vec{H}_1 \cdot \vec{\sigma}_1 + \vec{H}_2 \cdot \vec{\sigma}_2 + \vec{H}_3 \cdot \vec{\sigma}_3 + H_0'\sigma_1^z + \frac{1}{2}H_0'\sigma_2^z \\ -g(\sigma_1^x\sigma_2^x + \sigma_1^y\sigma_2^y) - g(\sigma_2^x\sigma_3^x + \sigma_2^y\sigma_3^y)\Big) \quad (6)$$

The topological phases measured in these systems can be mathematically connected via projection functions:

$$\tilde{g}'(\tilde{H}_r, \tilde{H}_0) = \frac{1}{2\tilde{H}_r}\sqrt{1-(1-|1-2\tilde{H}_0|)^2}, \tilde{H}_0'(\tilde{H}_r, \tilde{H}_0) = \frac{1}{\tilde{H}_r}(1-|1-2\tilde{H}_0|) \quad (7)$$

where $\tilde{g}' = g/H_r, \tilde{H}_0' = H_0/H_r$ and $\tilde{H}_r = H_r/A_{\parallel}, \tilde{H}_0 = H_0/A_{\parallel}$.

In Fig. 4b, first, we numerically calculated the topological phase diagram of the interacting three-qubit system. For a large $\tilde{H}_r$, where the Rabi frequency becomes significantly larger than $A_{\parallel}$, the normalized coupling strength $\tilde{g}'$ approaches 0, where the three qubits distinctively contribute to the total Chern number to be $C = 3$. Owing to the inverse relation, when $\tilde{H}_r$ approaches 0, both $\tilde{g}'$ and $\tilde{H}_0'$ become large, where $C = 0$ in the phase diagram. The Chern number trait for these limiting cases remains similar to that of the coupled two-qubit Hamiltonian[25]. Meanwhile, a more complex phase structure can be found by analytically calculating the positions of the ground state degeneracy points with respect to the sweep parameter sphere manifold. The white dashed boundaries clarify four distinctive regions where the Chern number in each region corresponds to the number of monopoles enclosed by the surface. Along the $\tilde{H}_0'$ axis, one monopole exits the surface at $\tilde{H}_0' = 1$ ($C = 3$ to $C = 2$) and second at $\tilde{H}_0' = 2$ ($C = 2$ to $C = 1$). Next, along the $\tilde{g}'$ axis, the two monopoles escape the surface at $\tilde{g}' = 1/\sqrt{2}$, inducing the Chern number transition from $C = 3$ to $C = 1$.

Finally, we project the three-monopole topological phase measurements onto the interacting three-qubit system using Eq. (7) and then compared with the three-qubit Chern number simulation results (Fig. 4c). For a fixed $\tilde{H}_0$, $\tilde{H}_r$ is swept from 0.22 to 2.2 by varying the Rabi frequency on the NV spin. The orthogonal parameter axes, $\tilde{H}_0$ and $\tilde{H}_r$, are nonlinearly transformed into $\tilde{H}_0'$ and $\tilde{g}'$ which gives topological phase transition curves in radial cross-sections for $\tilde{H}_0 = 0$, 0.23, 0.45, 0.68, and 0.91. The three-monopole Chern number transition projection, evaluated using Eq. (7), and the simulated Chern number transition cross section of the interacting three-qubit system are consistent with each other (blue dotted line in Fig. 4b).

The coupled multiqubit Hamiltonian carries multiple degenerate ground states, which leads to the realization of a high Chern number. Here, the interaction strength, $g$, between qubits determines the position of the monopoles on the parameter space z-axis (see Supplementary Information). In principle, investigating the topology of an N-interacting qubit system could simulate the topology of non-interacting 2N band models; for example, two interacting qubit systems simulating the topology of the ground band of this four-band electronic model and an interacting three-qubit system could help to probe the topological structure of the half-filled eight-band model[25].

Our scheme clearly shows that a high Chern number can be deterministically simulated using a single-qubit-based multi-monopole system, in addition to tuning the level of its transition depending on the range of $H_0/A_\parallel$ and $H_r/A_\parallel$ variations. For example, an electron-nuclear spin coupled system in diamond can be a versatile tool for studying a high-dimensional topology because further scaling up to a higher topological invariant number can be straightforwardly performed by utilizing the intrinsic $^{13}$C nuclear spins near the NV spin qubit with hyperfine coupling strengths varying from a few tens of kHz to almost ~100 MHz[36]. For a higher-number symmetric monopole system, one can engineer the Chern number transition with an increment of 1 or an even number transition: $C = 0,2,4 \cdots$ or an odd number transition $C = 1,3,5 \cdots$ by tuning the detuning $H_0/A_\parallel$.

**Conclusion**

In conclusion, we simulated a high topological invariant number using a simple system of NV electronic spin qubit hyperfine coupled with $^{14}$N nuclear spin and demonstrated the robust tunability of the measured topological invariant number up to $C = 3$ by harnessing the control parameters of the qubit. A systematic design of the Hamiltonian parameter sphere reveals the detailed topological structures over the three synthetic monopoles as well as the intriguing Chern number physics associated with the adiabaticity of the system's evolution over time. The generality of this method can be expanded to various qubit platforms to investigate the topology of higher dimensions, such as N-interacting qubit systems, which can simulate the topology of non-interacting 2N band models in condensed-matter physics. Furthermore, the tunability of the topological invariant of a qubit system can be directly applied to explore more exotic topology, which could be applied to the field of topological quantum information science.

**Methods:**

*NV spin system with three degeneracies*

The NV centre ground-state has an electronic spin with a spin quantum number of $S = 1$ with sublevels $|0\rangle$ and $|\pm1\rangle$. However, throughout this work, we only used $|0\rangle$ and $|-1\rangle$ as a two-level system by Zeeman splinting the $|\pm1\rangle$ states using a static external field. This electronic spin experiences a hyperfine interaction with the host nuclear spin $^{14}$N with a spin quantum number of $I = 1$. The longitudinal component of the hyperfine interaction with a coupling strength of $A_\parallel = 2.2$ MHz further splits the degeneracy of $|-1\rangle$ into three levels. The internal Hamiltonian assumes the form $\widehat{H}_0 = A_\parallel \sigma^z I^z$. Consequently, this electronic-nuclear spin system contains three degeneracy points, allowing us to simulate topological phases with a Chern number greater than 1. Additionally, the transition between different Chern numbers can be realized by introducing a common offset to these degeneracy points. The topology realized in this study corresponds to an eight-band noninteracting triangular lattice model.

*Experimental setup*

Measurements were performed using a home-built NV-diamond confocal microscope. An acousto-optic modulator (Isomet Corporation) enabled time gating of a 400 mW, 532 nm diode-pumped solid-state laser (Changchun New Industries). The laser beam was coupled to a single-mode fibre, and subsequently, delivered to an oil-immersion objective (100x, 1.3 NA, Nikon CFI Plan Fluor), and focused onto a diamond sample. The diamond sample was fixed on a three-axis motorized stage (Micos GmbH) for precise position control. NV red fluorescence (FL) was collected using the same objective and then passed through a dichroic filter (Semrock LP02-633RS-25). A pinhole (diameter 75 μm) was used with a f = 150 mm telescope to spatially filter the FL signal, which was detected using a silicon avalanche photodetector (Perkin Elmer SPCM-ARQH-12). A signal generator (SG, Agilent E4428C) was used to provide the carrier microwave signal. The phase and amplitude of the carrier signal was modulated with a 1 G/s rate arbitrary waveform generator (AWG, Tektronix AWG 5014C) and an IQ mixer (Marki IQ 1545 LMP). The microwave sideband signal was amplified (Mini-circuits ZHL-16W-43-S+) and passed through a gold coplanar waveguide, fabricated on a quartz coverslip using photolithography, and was mounted directly on the diamond sample to control the NV spin qubit. The diamond sample was CVD-grown, $^{12}$C isotopically purified to 99.99 %, and had dimensions of 2 mm × 2 mm × 0.5 mm. After the implantation of $^{14}$N$^+$ ion, the diamond was annealed at 800 °C for 8 h and at 1000 °C for 10 h. During the measurement, the external magnetic field was aligned with the NV crystalline axis with a field strength of ~100 G. The NV spin resonance lifetimes were $T_1 \sim 3$ ms, $T_2 \sim 400$ μs, and $T_2^* \sim 40$ μs.

*Quantum state tomography*

The general scheme of the control pulse to create a hemispherical trajectory is a sine enveloped chirped signal because of the sweeping of both the detuning and Rabi frequency. To match the relative phase of the chirped signal, we connected the tomography pulse directly after the control pulse at a given time. $T_{\text{meas}}$. $\langle \sigma^y \rangle$ rotation tomography pulse's relative phase was set with respect to the end phase of the chirped control signal. The tomography pulse Rabi frequency was set at 10 MHz. During the tomographic pulse calibration, we observed the dynamic phase noise contribution to be highly suppressed. Final NV spin state is readout by observing the amount of fluorescence in the 640-800 nm band, caused by an optical illumination of 532 nm. A change in fluorescence intensity occurs due to a non-radiative decay pathway via metastable singlet states (Fig. 1b). Because the NV spin qubit system has two isolated levels under an external bias magnetic field (~100 G), leakage to other states can be neglected, which gives the fidelity advantage of using an NV centre for quantum simulations.

*Adiabaticity parameter determination*

We determined an optimized condition for adiabaticity parameter $\alpha$, where the qubit response is quasi-adiabatic. This condition is fulfilled when the qubit adiabatically follows the Larmor vector trajectory yet the observable $\langle \sigma^y \rangle$, which is the Lorentzian deviation from the Larmor vector trajectory, has a sufficiently large signal-to-noise ratio to be detected. Using dynamic-state preparation as a benchmark of adiabaticity parameter calibration[25], we first detected the Landau-Zener transition (supplementary information) by measuring $\langle \sigma^z \rangle$ the hemispherical manipulation of a spin qubit and confirmed that the transition probability depends on the Rabi frequency $\Omega$ and $T_{\text{ramp}}$. The Landau-Zener $\langle \sigma_z \rangle$ measurements, obtained by varying $\alpha$, prove that our system's quasi-adiabatic boundary is approximately with the $2 \leq \alpha \leq 10$ range.

*Numerical simulations*

All numerical simulations of the NV spin evolution in this work were performed by computing the time-ordered time evolution operator at each time step.

$$U(t_i, t_f) = \hat{T}\left\{\exp\left(-i\int_{t_i}^{t_f} H(t)dt\right)\right\} = \prod_{j=1}^{N} \exp(-i\Delta t H(t_j)) \quad (7)$$

where $t_i$ and $t_f$ denote the initial and final time, respectively; $\hat{T}$ is the time-ordering operator; $\Delta t$ is the time step size of the simulation; $N = (t_f - t_i)/\Delta t$ is the number of time steps; and $\hat{H}(t)$ is the time-dependent Hamiltonian. In the simulation, we used a step size of $\Delta t = 1$ ns, which was sufficiently small in the rotating frame. The algorithm was implemented using the MATLAB® software.

**Data availability**

The data supporting the findings of this study are available from the corresponding author upon reasonable request.

**Code availability**

The codes used in this study are available from the corresponding author upon reasonable request.

**Acknowledgements**

The authors would like to thank A. Polkovnikov and M. Kolodrubetz for their fruitful discussions. This material is based upon work supported by, or in part by, the U.S. Army Research Laboratory and the U.S. Army Research Office under contract/grant numbers W911NF1510548 and W911NF1110400. This work was performed in part at the Center for Nanoscale Systems (CNS), a member of the National Nanotechnology Coordinated Infra-structure Network (NNCI), which is supported by the National Science Foundation under NSF award no. 15419. J. L. was partially supported by the ILJU Graduate Fellowship and the KIST research program (2E31531). KA received funding from the JST PRESTO (Grant Number JPMJPR20B1).

**Author contributions**

The project was conceived by J. L. and K. A., and supervised by R.W. J. L., K. A., and H. Z. designed the experimental set-up. J. L. and K. A. performed the measurements and analysed the data. All authors contributed to the discussion and preparation of the manuscript.

**Competing interests**

The authors declare no competing interests.

**Additional information**:

Supplementary Information is available for this paper.

Correspondence and requests for materials should be addressed to walsworth@umd.edu.

**Figures**

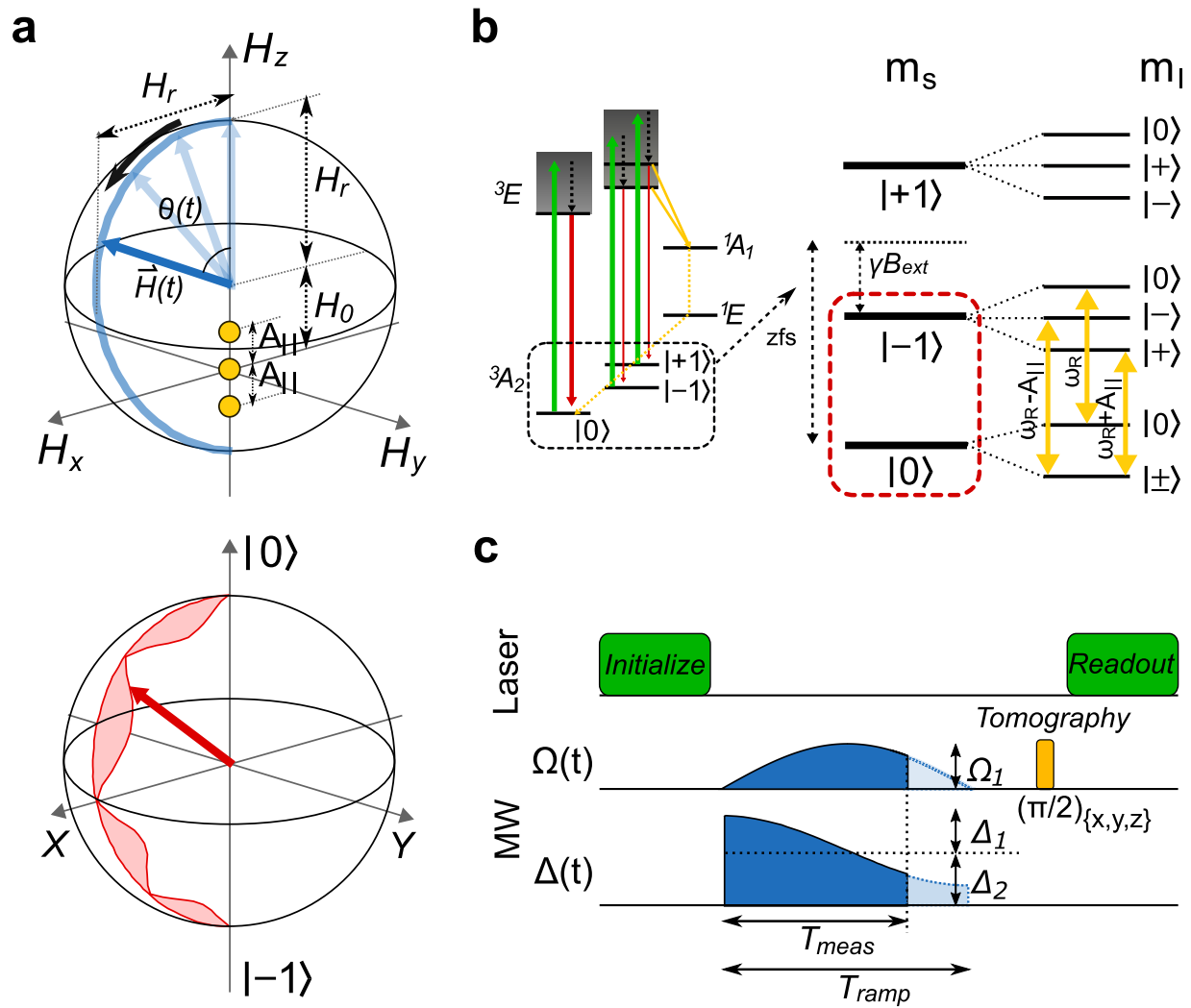

Figure 1. Schematic of the Chern number measurement approach using an electronic-nuclear spin system associated with an NV centre in diamond. **a,** (top) Trajectory of the Larmor

vector $\vec{H}(t)$, represented by a thick blue arrow, in the control Hamiltonian parameter space $(H_x, H_y, H_z)$. The solid black arrow indicates the direction of the sweep of the Larmor vector. The yellow circles represent the degeneracy points of the system Hamiltonian. (bottom). Time evolution of the Bloch vector $\vec{\sigma}(t)$ is represented by the red arrow on the Bloch sphere. The Bloch vector path deviates from the Larmor vector path due to a nonadiabatic response. The red filled area indicates the amount of deviation, a summation of which over the path is related to the Chern number. **b,** NV centre energy level diagram. The NV ground states consist of $|0\rangle, |\pm1\rangle$ electronic spin sublevels, which are further split by hyperfine interactions with the host $^{14}$N nuclear spin. Three hyperfine transitions between $|0\rangle$ and $|-1\rangle$ electronic spin sublevels (yellow double-arrows) define three degeneracy points in the rotating frame with angular speed of $\omega_R - A_\parallel, \omega_R, \omega_R + A_\parallel$, where $A_\parallel$ is the parallel component of the hyperfine tensor. **d,** Experimental pulse sequence. An initialization pulse polarizes the electronic spin into $|0\rangle$. Then, a microwave pulse with time-varying Rabi frequency $\Omega(t) = \Omega_1 \sin\theta(t)$ and time-varying detuning $\Delta(t) = \Delta_1 \cos\theta(t) + \Delta_2$ under a constraint of $\Omega_1 = \Delta_1$ realizes the Larmor vector trajectory. This pulse is terminated at $t = T_{meas}$. The Larmor vector trajectory is completed within a time of $T_{ramp}$. The combination of a microwave tomography pulse and a laser readout pulse allows the measurement of all the Bloch vector components.

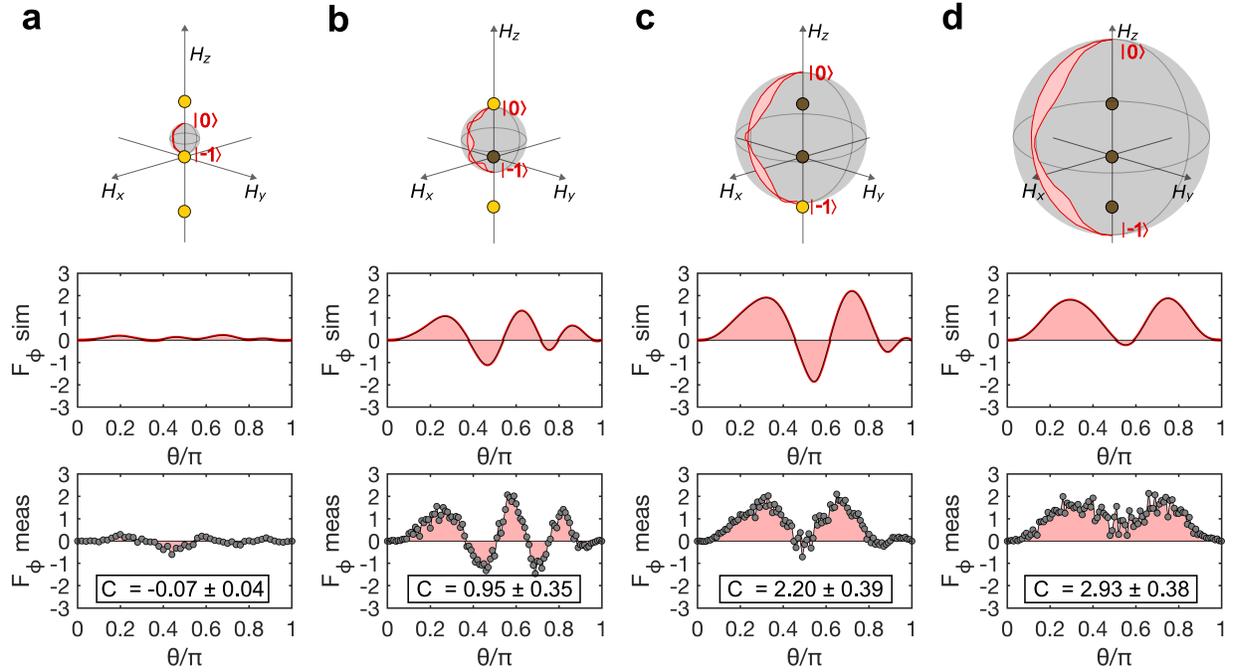

**Figure 2. Observation of Chern numbers from 0 to 3.** The number of degeneracy points included in the sphere are incremented one by one by enlarging the normalized radius $H_r/A_\parallel$ under a fixed offset of $H_0/A_\parallel = 0.23$. **a,** Case with no degeneracy point. Expected Chern number is $C = 0$. (Top) Illustration of the Bloch vector trajectory (red line) overlayed with the degeneracy points (yellow and grey circles). (Middle) Numerically simulated Berry curvature $F_\phi^{\text{sim}}(\theta)$ (red line). (Bottom) Measured Berry curvature $F_\phi^{\text{meas}}(\theta)$ (grey circles and red shaded area). Integrating the Berry curvature over $\theta$ gives a Chern number of $C = -0.07 \pm 0.04$. **b-d,** Cases with one, two, and three degeneracy points, respectively. Expected Chern numbers are $C = 1, 2, 3$, while the measured Chern numbers are $C = 0.95 \pm 0.35, 2.20 \pm 0.39, 2.93 \pm 0.38$.

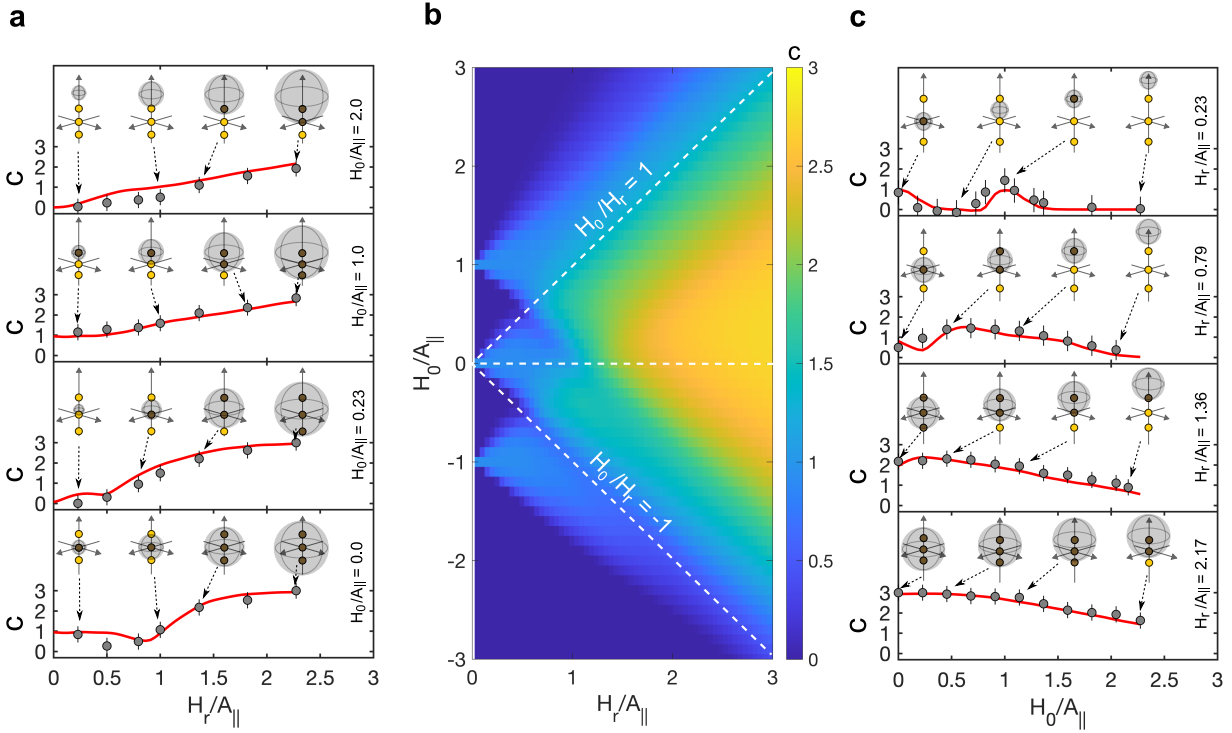

**Figure 3. Two-dimensional topological phase diagram and phase transitions along the vertical and horizontal directions. a,** Measured (grey circles) and simulated (red solid line) Chern number values along the normalized radius $H_r/A_\parallel$ direction under a fixed normalized offset of $H_0/A_\parallel = 2.00, 1.00, 0.23, 0.00$. **b,** Numerically simulated topological phase map. White dashed lines along $H_0/H_r = \pm 1$ are presented as a guide for eye. Asymmetry in the pattern with respect to $H_0/A_\parallel = 0$ can be attributed to the time-reversal symmetry breaking of our measurement protocol. **c,** Measured (grey circles) and simulated (red solid line) Chern number values along the normalized offset $H_0/A_\parallel$ for a fixed normalized radius of $H_r/A_\parallel = 0.23, 0.79, 1.36, 2.17$.

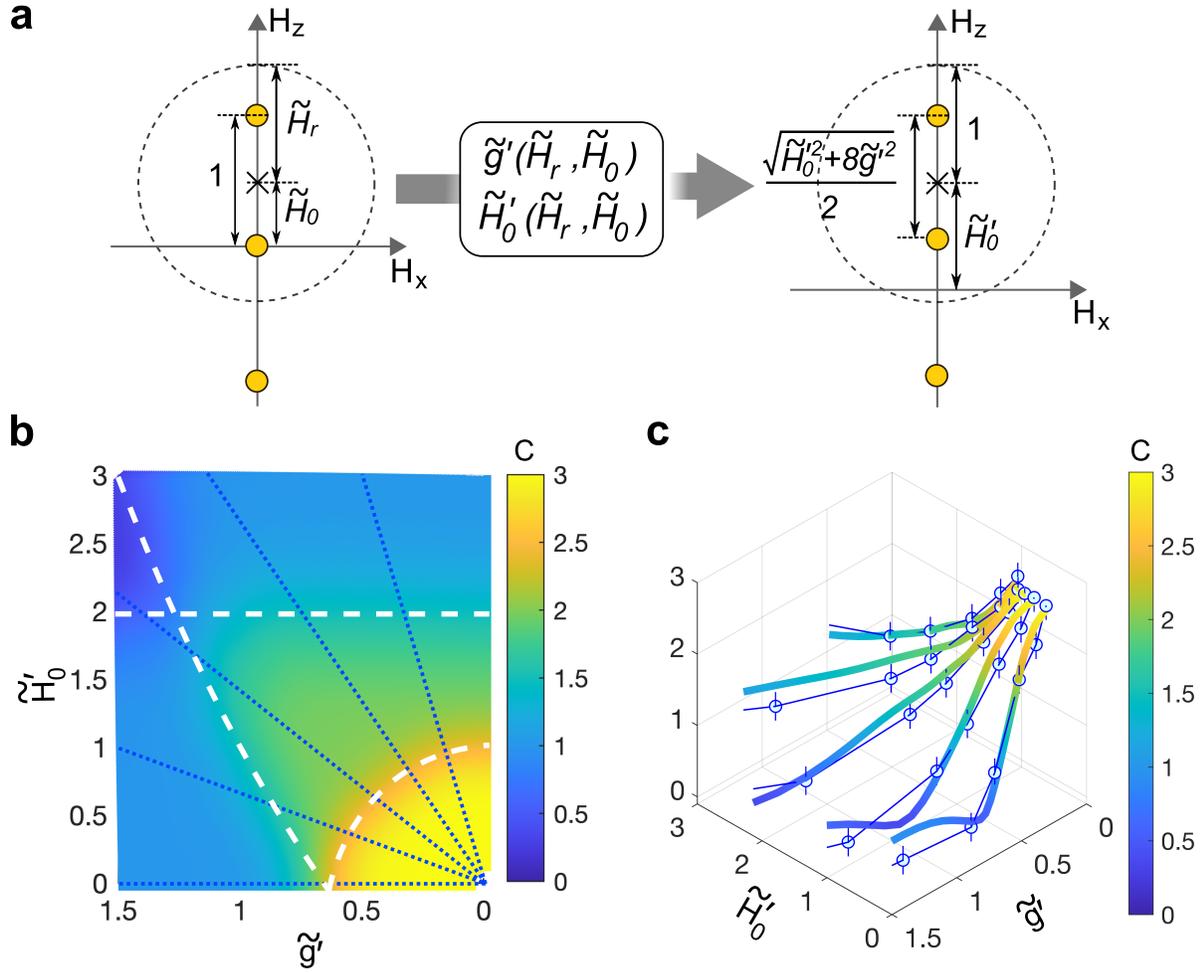

**Figure 4. Tuneable topological invariant with three monopoles. a**, Topological equivalence of the Chern number measurement using an electronic-nuclear spin system of the NV centre and three interacting qubits. (left) Degeneracy points of the NV electronic-nuclear spin system. The inter-degeneracy spacing $A_\parallel$ is constant, whereas, the radius $H_r$ and offset $H_0$ are variable. (right) Degeneracy points of the equispaced interacting three-qubit system. The inter-qubit coupling strength $g$ and the radius $H_r'$ are variable, while the offset $H_0'$ is fixed. **b**, Simulated Chern number phase diagram for the interacting three-qubit Hamiltonian. White dash lines indicate the transition boundaries, and the blue dotted lines denote the radial cross-sections presented in **c**. Projected three-monopole Chern number measurements (blue open circle) and the cross-sectional transition of a simulated three-qubit phase diagram (parula). Sweep parameters are normalized for attaining a unitless topology (three-monopole system normalized by $A_\parallel$ and three-qubit system normalized by $H_r'$). Each radial projection corresponds to a fixed $\tilde{H}_0 = 0, 0.23, 0.45, 0.68$ and $0.91$.